\documentclass[sigconf,screen]{acmart}

\usepackage{url}
\usepackage{soul}
\usepackage{textcomp}
\usepackage{stfloats}
\usepackage{verbatim}
\usepackage{multirow}
\usepackage{siunitx}
\usepackage{makecell}
\usepackage{colortbl}
\usepackage{enumitem}
\usepackage{amsfonts}
\usepackage{algorithmic}
\usepackage{fontawesome}
\usepackage[caption=false,font=normalsize,labelfont=sf,textfont=sf]{subfig}
\usepackage{microtype}

\AtBeginDocument{
  \providecommand\BibTeX{{
    \normalfont B\kern-0.5em{\scshape i\kern-0.25em b}\kern-0.8em\TeX}}}

\definecolor{mygray}{gray}{.9} 

\newcommand{\techtitle}{WiseSpec}
\newcommand{\tech}{\textit{\techtitle{}}}

\newcommand{\Comment}[1]{}

\NewDocumentCommand{\framecolorbox}{oommm}
 {
  \IfValueTF{#1}
   {\IfValueTF{#2}
    {\fcolorbox{#3}{#4}{\makebox[#1][#2]{#5}}}
    {\fcolorbox{#3}{#4}{\makebox[#1]{#5}}}%
   }
   {\fcolorbox{#3}{#4}{#5}}%
 }

\setcopyright{cc}
\setcctype{by-nc-nd}
\acmDOI{10.1145/3832783.3844573}
\acmYear{2026}
\copyrightyear{2026}
\acmISBN{979-8-4007-2882-2/2026/10}
\acmConference[ASE '26]{Proceedings of the 41st IEEE/ACM International Conference on Automated Software Engineering}{October 12--16, 2026}{Munich, Germany}
\acmBooktitle{Proceedings of the 41st IEEE/ACM International Conference on Automated Software Engineering (ASE '26), October 12--16, 2026, Munich, Germany}
\acmSubmissionID{ase26src-p58-p}
\received{2026-07-12}
\received[accepted]{2026-08-25}

\begin{document}

\title{\techtitle{}: Requirements-Driven Agents for Code Generation}

\author{Zhao Tian}
\orcid{0000-0002-9316-7250}
\affiliation{
  \institution{School of Computer Software, Tianjin University}
  \city{Tianjin}
  \country{China}
}
\email{tianzhao@tju.edu.cn}

\begin{abstract}
Code generation aims to automatically generate source code from task requirements and has attracted significant attention with the rapid advancement of large language models (LLMs). 
Despite remarkable progress, LLMs often struggle to generate correct code for complex software engineering tasks because task descriptions are frequently incomplete, ambiguous, or lack critical contextual information. 
Existing approaches primarily improve the capabilities of coding agents through more sophisticated tools, skills, and workflows, while largely overlooking the quality of the task requirements themselves. 
To address this limitation, we draw inspiration from software requirements engineering and propose \tech{}, a novel requirements-driven agent framework for repository-level code generation. 
\tech{} automatically constructs structured and information-rich requirements, assesses their quality through execution-based evaluation, and iteratively refines them to better guide code generation. 
Experimental results show that \tech{} consistently outperforms all baselines, achieving an average improvement of 13.17\% in \textit{\%Resolved}.
\end{abstract}

\ccsdesc[500]{Software and its engineering~Automatic programming}

\keywords{Code Generation, Agent, Requirements Engineering}

\maketitle

\section{Introduction}
\label{sec:introduction}

Code generation aims to automatically generate source code from programming requirements, offering substantial potential to improve developer productivity and software quality~\cite{mastropaolo2023robustness,becker2026can}.
Recent advances in Large Language Model (LLM)-based coding agents have achieved remarkable progress by enhancing the capabilities of LLMs through sophisticated tools, skills, and workflows~\cite{yang2024swe,gao2025trae}. 
Despite these advances, complex software engineering tasks remain challenging. 
Existing approaches primarily focus on improving how LLMs solve programming tasks, while largely overlooking what they are asked to solve, namely the quality of task requirements themselves~\cite{kuang2026reagent}. 
Most methods directly consume the original problem description as input, implicitly assuming that it faithfully specifies the intended program behavior. 
In practice, however, task descriptions are frequently ambiguous, incomplete, or missing critical contextual information, making them an unreliable representation of the underlying requirements~\cite{tian2025fixing,tian2025aligning}. 
Consequently, even powerful LLMs struggle to accurately infer user intent and produce correct implementations, highlighting the need for explicit requirement understanding and alignment before code generation.

To address these challenges, we propose \textbf{\tech{}}, a novel requirements-driven agent that enhances the code generation performance of LLMs. 
First, \tech{} collects relevant contextual information to construct structured, information-rich requirements using a predefined domain-specific language (DSL). 
Second, it reformulates requirement quality assessment as an execution-based code evaluation problem, enabling the computation of a quantitative requirement quality score. 
Third, \tech{} iteratively refines and aligns the generated requirements according to refinement and alignment rules, ultimately producing higher-quality requirements for better code generation.
Experiment results demonstrate that \tech{} significantly outperforms all three state-of-the-art baselines across two LLMs and three benchmarks.
\begin{figure}[t]
    \centering
    \includegraphics[width=1.0\linewidth]{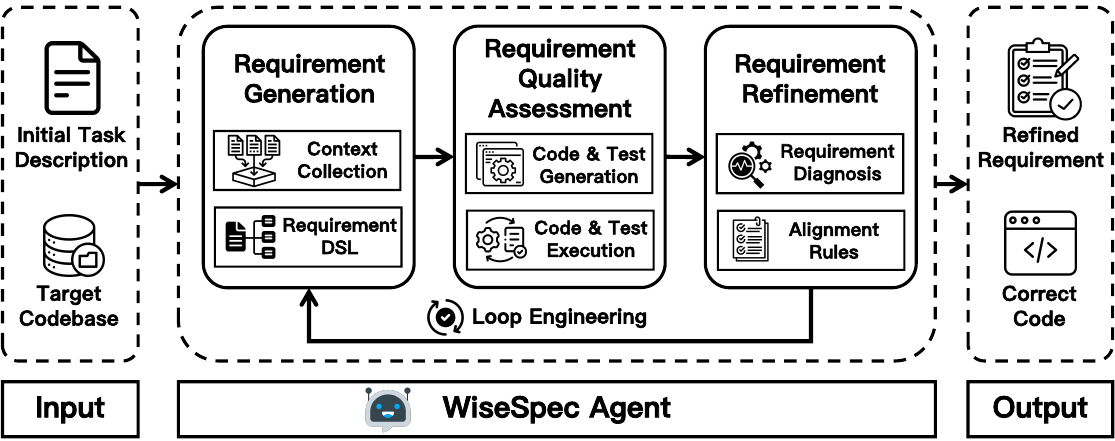}
    \caption{The overview of \tech{}}
    \label{fig:overview}
    \vspace{-6mm}
\end{figure}

\section{Approach}
\label{sec:approach}
Figure~\ref{fig:overview} illustrates the overview of \tech{}, consisting of three components: Requirement Generation, Requirement Quality Assessment, and Requirement Refinement.

\subsection{Requirement Generation}
To accurately retrieve the contextual information required for code generation, \tech{} simulates the program comprehension process by iteratively collecting and analyzing relevant code snippets from the target codebase. 
Starting from the given task description, it progressively expands the retrieval scope along program dependencies, using previously collected code snippets to guide subsequent retrieval and exploration. 
To transform the collected yet fragmented information into structured programming requirements, \tech{} employs a predefined requirement DSL. 
The requirement DSL consists of nine primary requirement attributes and seventeen corresponding sub-attributes, covering both high-level architectural information and fine-grained implementation details.
Based on requirement DSL, \tech{} systematically organizes the retrieved contextual information into a structured and information-rich requirement representation, providing a solid foundation for subsequent code generation.

\subsection{Requirement Quality Assessment}
Assessing the quality of requirement specifications is a fundamental step in requirements engineering, as ambiguous, incomplete, or incorrect requirements can propagate errors to downstream implementations~\cite{han2024archcode}. 
However, because requirements are typically expressed in structured natural language without formal semantics, their quality is difficult to evaluate directly and quantitatively~\cite{montgomery2022empirical}. 
To address this challenge, \tech{} reformulates requirement quality assessment as an execution-based code evaluation problem. 
Specifically, it first generates executable code and tests from the synthesized requirements and evaluates the generated code through test execution. 
Since code is expected to faithfully implement the intended requirements, its execution correctness serves as an effective proxy for requirement quality. 
\tech{} adopts a strict acceptance criterion, i.e., the generated code is accepted only if it passes all generated tests. 
Otherwise, the corresponding requirements are considered potentially deficient and are forwarded to the subsequent requirement refinement and alignment stage.

\subsection{Requirement Refinement}
This component iteratively improves low-quality requirements to better guide LLMs toward generating correct code. 
To diagnose requirement deficiencies, \tech{} categorizes them into three mutually exclusive and collectively exhaustive types: \textit{Conflict}, \textit{Omission}, and \textit{Ambiguity}. 
Based on the identified deficiency type, \tech{} applies a set of predefined requirement alignment rules to generate actionable refinement feedback. 
Guided by this refinement feedback, the requirements are iteratively refined and re-assessed.
During the refinement process, \tech{} adopts a greedy optimization strategy that retains the candidate requirement with the highest quality score at each iteration, as it is more likely to provide accurate and complete guidance for code generation. 
Furthermore, refinement feedback that fails to improve requirement quality is recorded as counterexamples, enabling \tech{} to adjust its refinement strategy in subsequent iterations. 
Through this loop engineering, \tech{} progressively improves the requirement quality, leading to more reliable code generation.

\section{Experiments and Results}
\label{sec:experiments_and_results}

~\indent
\ul{\textbf{I. Process:}}
To comprehensively evaluate \tech{}, we compare it against three state-of-the-art coding agents: Agentless~\cite{xia2025demystifying}, Trae-agent~\cite{gao2025trae}, and Claude Code~\cite{anthropic2026claudecode}. 
The evaluation is conducted on three widely used repository-level code generation benchmarks: SWE-bench-Lite~\cite{jimenez2024swe}, SWE-bench-Verified~\cite{openai2024swebenchverified}, and SWE-bench-Pro~\cite{deng2025swe}. 
We randomly sample 100 instances from each benchmark to control the computational cost. 
We use two advanced LLMs, DeepSeek-V3.2~\cite{liu2025deepseek} and Qwen-Plus-2025-12-01~\cite{yang2025qwen3}, as the underlying models. 
We evaluate all approaches using two metrics: \textit{\%Applied}, which measures the syntactic correctness of generated code by determining whether it can be successfully applied to the codebase, and \textit{\%Resolved}, which measures functional correctness by assessing whether the generated code passes all gold tests.

\ul{\textbf{II. Results:}}
Table~\ref{tab:effectiveness} presents the effectiveness comparison of all approaches. 
Across all six experimental settings (3 benchmarks $\times$ 2 LLMs), \tech{} consistently achieves the best performance, outperforming all representative coding agents. 
Specifically, \tech{} improves \textit{\%Resolved} by 2\%$\sim$29\% and \textit{\%Applied} by 11\%$\sim$63\% over the  baselines across different settings. 
To further evaluate its generalizability to more capable LLMs, we conduct an additional experiment using the state-of-the-art Claude-Opus-4.8 on SWE-bench-Pro. 
While Claude Code achieves a \textit{\%Resolved} score of 53\%, \tech{} further increases the score to 56\%, demonstrating that it generalizes well to stronger foundation models. 
Overall, these results show that the proposed requirements-driven paradigm effectively improves repository-level code generation. 
Furthermore, a Wilcoxon signed-rank test ($\alpha=0.05$) yields $p<2.5\times10^{-4}$, confirming that the improvements of \tech{} over all baselines are statistically significant for both \textit{\%Resolved} and \textit{\%Applied}.

\begin{table}[t!]
    \caption{Comparison in \textit{\%Applied} (\textcolor{green}{$\uparrow$}) and \textit{\%Resolved} (\textcolor{green}{$\uparrow$}).}
    \vspace{-3mm}
    \label{tab:effectiveness} 
    \centering
    \setlength{\tabcolsep}{1.6mm}
    \normalsize
    \begin{tabular}{lcccccc}
    \toprule
    \multirow{2}{*}{\textbf{Technique}}
    & \multicolumn{2}{c}{\textbf{SWE-Lite}}
    & \multicolumn{2}{c}{\textbf{SWE-Verified}}
    & \multicolumn{2}{c}{\textbf{SWE-Pro}} \\
    \cmidrule(lr){2-3} \cmidrule(lr){4-5} \cmidrule(lr){6-7}
    & \%App. & \%Res.
    & \%App. & \%Res.
    & \%App. & \%Res. \\
    \midrule
    \multicolumn{7}{l}{\cellcolor{gray!30}\textbf{DeepSeek}} \\
    Agentless   & 55\% & 24\% & 61\% & 35\% & 62\% & 6\% \\
    Trae-agent  & 64\% & 28\% & 61\% & 35\% & 43\% & 11\% \\
    Claude Code & 72\% & 36\% & 77\% & 46\% & 84\% & 24\% \\
    \textbf{\tech{}} & \cellcolor{green!20}\textbf{100\%} & \cellcolor{green!20}\textbf{39\%} & \cellcolor{green!20}\textbf{93\%} & \cellcolor{green!20}\textbf{51\%} & \cellcolor{green!20}\textbf{100\%} & \cellcolor{green!20}\textbf{35\%} \\
    \midrule
    \multicolumn{7}{l}{\cellcolor{gray!30}\textbf{Qwen}} \\
    Agentless   & 47\% & 14\% & 35\% & 22\% & 59\% & 5\% \\
    Trae-agent  & 55\% & 17\% & 63\% & 24\% & 49\% & 5\% \\
    Claude Code & 89\% & 26\% & 80\% & 33\% & 77\% & 20\% \\
    \textbf{\tech{}} & \cellcolor{green!20}\textbf{100\%} & \cellcolor{green!20}\textbf{28\%} & \cellcolor{green!20}\textbf{98\%} & \cellcolor{green!20}\textbf{37\%} & \cellcolor{green!20}\textbf{99\%} & \cellcolor{green!20}\textbf{26\%} \\
    \bottomrule
    \end{tabular}
    \vspace{-7mm}
\end{table}

\section{Conclusion}
\label{sec:conclusion}
In this paper, we identify the quality of task requirements as a fundamental bottleneck in repository-level code generation and propose a requirements-driven paradigm to address this challenge. 
Based on this insight, we present \tech{}, a novel requirements-driven agent that automatically constructs structured and information-rich requirements, assesses their quality through execution-based evaluation, and iteratively refines them to improve code correctness. 
Experiment results demonstrate that \tech{} consistently outperforms all baselines across multiple evaluation metrics, highlighting the effectiveness of requirements engineering in improving LLM-based code generation.


\begin{acks}
Zhao Tian is advised by \textbf{Professor. Junjie Chen}.
This work is supported by National Natural Science Foundation of China (Grant No. 62322208).
\end{acks}

\bibliographystyle{ACM-Reference-Format}
\bibliography{reference}

\end{document}